# Cerebro spinal fluid dynamic in front of cardiac and breathing influence

Olivier Baledent[1], Pan Liu[1], Serge Metanbou[1], Cyrille Capel[1], Sidy Fall[1], and Roger Bouzerar[1]

[1]university hospital Jules Verne, Amiens, France

## Synopsis

it is still debated how breathing interact with the CSF. New Phase contrast MRI sequence based on Echo Planar imaging (EPI-PC) can now produce continuously during minutes a velocity map, more or less every 100 ms. We did not found in the literature quantitative evaluation of the CSF stroke volume change during breathing. The aim of this work is to quantify CSF dynamics change in the aqueduct and in the spinal canal during the breathing and cardiac period using EPI-PC.

## Background

Conventional cine Phase contrast MRI (Conv-PC) can quantify Cerebro Spinal Fluid (CSF) dynamic by using cardiac synchronization (1). This sequence can reconstruct only one mean cardiac flow curve from all the cardiac cycles of the acquisition, acquisition that can take minutes. Since the 1990ies many works have demonstrate how the CSF dynamics interact with the arterial and venous flows (2-4). Nevertheless it is still debated how breathing interact with the CSF. Some authors have shown that breathing influence more or less the CSF dynamics (5-8). Some of them postulate that CSF does not change its velocities direction during many cardiac cycles but only by the influence of breathing. Whereas others continue to think that in normal condition, CSF goes up and down in respons of the systolic and diastolic period of the cardiac cycle. CSF stroke volume (SV) is a useful parameter well known by the persons working in this field that represents the CSF volume moving up down during the cardiac cycle. New Phase contrast MRI sequence based on Echo Planar imaging (EPI-PC) can now produce continuously during minutes a velocity map, more or less every 100 ms function of the machine and of the quality wanted (9). We did not found in the literature quantitative evaluation of the CSF stroke volume change during breathing. The aim of this work is to quantify CSF dynamics change in the aqueduct and in the spinal canal during the breathing and cardiac period using EPI-PC.

## Methods

The study was performed on a 3T scanner with a 32 channels head coil was used. Ten healthy adults volunteers signed the ethical agreement consent (ID-RCB : 2019-A02130-57) to investigate CSF flows in the aqueduct and at the cervical level (fig 1). Two different acquisitions were used (fig 2):
· Conventional 2D Cine-PC (Conv PC) using retrospective plethysmograph gating.
· 2D EPI PCMRI sequence free of any synchronization.
Post processing was done by homemade software to calculate CSF flow curves during cardiac cycle. Conv-PC provided only one flow curve to represent all the cardiac cycles of the acquisition whereas EPI-PC provided continuous flow dynamics curves including all the cardiac cycles of the acquisition. Then it was possible to study the change of CSF flows between inspiration and expiration. Each CSF flows curve obtained from EPI PC signals (8-12 points) per cardiac cycle was extrapolate to 32 points to be compared with conv PC.
Respiratory physiological signal was recorded during the MRI acquisition using a pneumatic belt sensor. This respiratory signal was used in the post processing to divide the inspiration and expiration periods and reconstruct CSF flows dynamic of these two different breathing periods. Then it was possible to study the change of CSF flows between inspiration and expiration. SV was calculated for conv PC and EPI PC and compared together.
Statistical spearman correlation and paired t-wilcoxon tests were used to compared EPI-PC and Conv-PC results.

## Results

Both for Conv PC and EPI PC, CSF flows were present and simply identifiable in the spinal spaces and in the aqueduct in all the PCMRI of the subjects. SV of the CSF in the aqueduct measured in the Conv PC and EPI PC were well correlated (Rs=0.94, p=0.0001) but the SV of CSF in the aqueduct measured by EPI PC were higher than those measured by Conv PC (p=0.009). SV of the CSF in the spinal canal measured in the Conv PC and EPI PC were little less correlated (Rs=0.67, p=0.039) and the SV of CSF in the spinal canal measured by EPI PC were this time smaller than those measured by Conv PC (p=0.032) (fig 3). An example of the Curves obtained by EPI PC is presented in fig (4-5). For all the subjects it was possible to calculate the SV of the CSF during inspiration and expiration periods. In the aqueduct and in the spinal spaces CSF SV were respectively 9% and 8% higher during the inspiration periods in comparison with the expiration periods. In all the subjects, even if exist a small difference in the SV during breathing, CSF presented all times a change in the velocities direction during cardiac cycle corresponding to existence of a flush period during cardiac systole and a cranium filling period during diastole.

## Discussion-Conclusion

Without any cardiac or respiratory gating, EPI PC offers possibility to explore continuously for the first time the CSF flows during many cardiac and respiratory cycles. Even if the spatial resolution is still limited these results shows a good correlation with the conv PC, actually the gold standard. The good thing it is that EPI PC was well correlated with the conv PC but the strange thing was that in the small aqueduct, EPI overestimate the SV in comparison with conv PC, whereas in the large spinal spaces the SV was under estimated. Nevertheless, in conclusion we have shown that physiological breathing modulate only the CSF flows signals by around 10% whereas the cardiac power is the main force able to change the direction of the flow during cardiac cycle.

## Acknowledgements

Thanks to the MRI technicians: Garance Arbeaumont -Trocmé - Julien Van Gysel and Héléna Freulet

Thanks to the French Research agency: ANR-FIGURES & Hanuman

Thanks Région Haut de France

Thanks to MRI Research GIE-FF and CHU Amiens Picardie

Thanks to all the volunteers who trust us and let us working with their spins.

Thanks to David Chechin for his scientific support.

## Figures

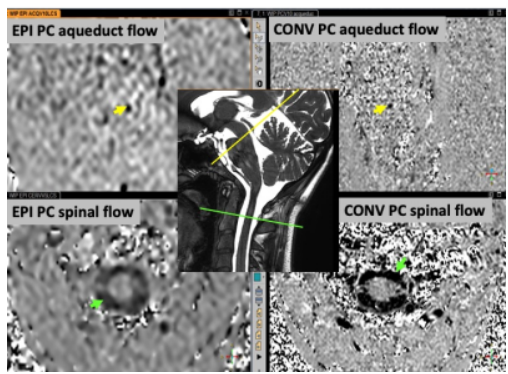

CSF flows in the aqueduct and spinal spaces.

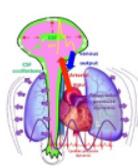

EPI PC and Conv PC protocol parameters

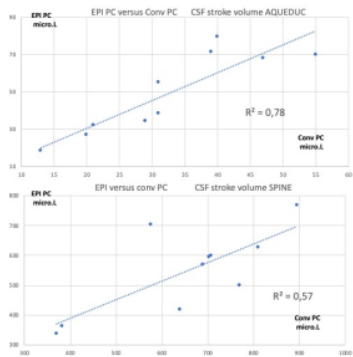

CSF Stroke volume : EPI PC versus conv PC

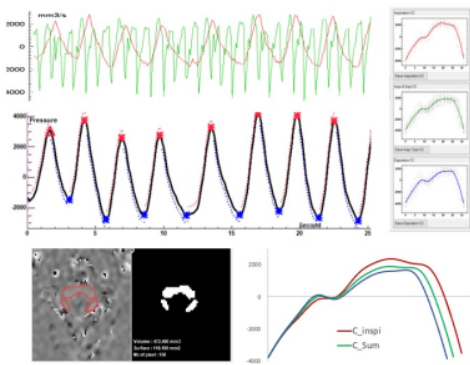

CSF flows continously acquired by EPI PC at the spinal level.

Base on the respiratory signal, EPI PC signal can be post processed. CSF flow dynamic is reconstructed along cardiac cyle during inspiration (red) and expiration (blue) periods. CSF flows oscillate around the zero line during each crdiac cycle.

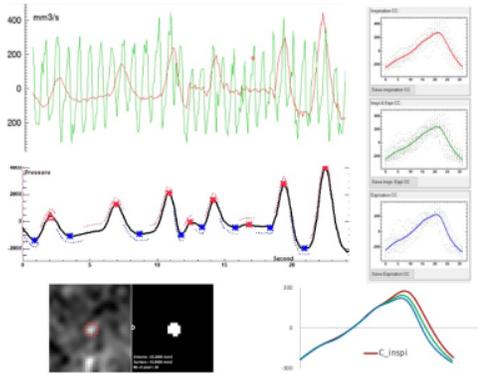

CSF flows continously acquired by EPI PC at the aqueductal level.

Base on the respiratory signal, EPI PC signal can be post processed. CSF flow dynamic is reconstructed along cardiac cyle during inspiration (red) and expiration (blue) periods. CSF flows oscillate around the zero line during each crdiac cycle.